\documentstyle[12pt]{article}
\begin{document}
%{\large
\bibliographystyle{unsrt}
\def\Journal#1#2#3#4{{#1}{\bf #2} (#4) #3}
\def\NCA{ Nuovo Cimento}
\def\NIM{ Nucl. Instrum. Methods}
\def\NIMA{{ Nucl. Instrum. Methods} A}
\def\NPB{{ Nucl. Phys.} {\bf B}}
\def\NPA{{ Nucl. Phys.} {\bf A}}
\def\PLB{{ Phys. Lett.}  {\bf B}}
\def\PRL{ Phys. Rev. Lett. }
\def\PRD{{ Phys. Rev.} {\bf D}}
\def\ZPC{{Z. Phys.} {\bf C}}
\def\ZPA{{Z. Phys.} {\bf A}}
\def\be{\begin{equation}}
\def\ee{\end{equation}}
\def\bea{\begin{eqnarray}}
\def\eea{\end{eqnarray}}

%\begin{flushright}
%AS-ITP-97-25 \\
%\today
%\end{flushright}

{\large\bf Twist-3 contribution to the pion electromagnetic form factor}

\vskip 1cm
\centerline{Fu-Guang Cao, Yuan-Ben Dai and Chao-Shang Huang}

\vskip 0.5cm
\centerline{
 CCAST (World Laboratory), P.O. Box 8730, Beijing 100080, P.~R.~China}

\centerline{and}

\centerline{
Institute of Theoretical Physics, Academia Sinica,
P.O. Box 2735, Beijing 100080, P.~R.~China\footnote{Mailing address.
E-mail: caofg@itp.ac.cn.}}

\vskip 2cm
\begin{abstract}
Non-leading contribution to the pion electromagnetic form factor
which comes from the pion twist-3 wave function is analyzed in the modified
hard scattering approach (MHSA) proposed by Li and Sterman.
This contribution is enhanced significantly due to bound state effect
(the twist-3 wave function is independent of the
fractional momentum carried by the parton and has a large factor
$\sim m_\pi^2/m_0$ with $m_\pi$ being the pion meson mass and $m_0$ being the
mean u- and d-quark masses).
Consequently, although it is suppressed by the factor $1/Q^2$,
%as compared with the leading (twist-2) contribution,
the twist-3 contribution is comparable with
and even larger than
the leading twist (twist-2) contribution at intermediate energy region
of $Q^2$ being $2 \sim 40 \,{\rm GeV}^2$.
%The perturbative predictions including both twist-2 and twist-3
%contributions are larger than the experiment data at lower
%energy regions which indicate that the other nonleading corrections
%may be important at these energy regions.

\bigskip
\bigskip
%\noindent
%PACS number(s): 12.38.Bx, 12.39.Ki, 13.40.Gp, 14.40.Aq

\end{abstract}

\newpage
%\section{Introduction}
\noindent
{\bf 1 Introduction}
\vskip 5pt

\noindent
There has been a lot of discussions about applying perturbative QCD (pQCD)
to exclusive processes at large momentum transfer [1-15].
Although there is general agreement that pQCD is able to make successful
predictions for
the exclusive processes at asymptotic limit $(Q^2\rightarrow \infty)$,
the applicability of pQCD to these processes at experimentally available
$Q^2$ region has been being debated and attracted much of attention.
The difficulties in practical calculation mainly
come from the end-point singularity, {\it i.e.} in the end-point
region ($x\rightarrow 0,1$ with $x$ being the fractional momentum carried by 
the parton) the virtuality of intermediate states is small and
the running couple constant $\alpha_s$ becomes large, 
thereby perturbation expansion might be illegal.
However, perturbative calculation can be rescued with the help of 
some techniques to cure the end-point singularity [8-15], for example,
the incorporation of the transverse structure of the pion wave 
function \cite{HuangS,DzieM,CaoGamma}, 
the introduction of an effective gluon mass \cite{JiSLNA}
and a frozen running coupling constant \cite{JiSLNA,Cornwall}.
%It is pointed out in \cite{HuangS} that the applicability of
%pQCD to the hadronic form factors is questionable only as momentum transfers
%being $Q^2\le 4$GeV$^2$.
Recently, Li and Sterman \cite{LiPion,LiProton}
proposed a modified hard scattering approach (MHSA)
for the hadronic form factor by taking into account the customarily 
neglected partonic transverse momentum as well as Sudakov corrections.
They point out that pQCD calculation for the pion form factor
begins to be self-consistent at about
$Q\sim 20 \Lambda_{QCD}$, which is similar to the conclusion given in
Ref. \cite{HuangS}.
%the applicability of pQCD to the hadronic form factors
%is questionable only as momentum transfers being $Q^2\le 4$GeV$^2$.
More recently, Ji, Pang and Szczepaniak \cite{Ji95} arrived at a similar
conclusion as Refs. \cite{HuangS,LiPion,LiProton} by analyzing
the factorization perturbation formalism for the pion form factor 
in the framework of light-cone time-order perturbative theory.
These studies shed light on applying pQCD to exclusive
processes at intermediate energy region.

However, there is still a crucial problem which has not been solved,
%the above arguments have not address,
that is although improved pQCD calculation for the exclusive processes
is self-consistent at currently experimentally accessible energy region,
the numerical predictions are generally far smaller than the experimental data.
For example, pQCD prediction for the pion form factor is
\bea
F_\pi (Q^2\rightarrow\infty)
&=&16\pi \alpha_s(Q^2) C_F
\int[dx][dy] \phi(x) \frac{1}{x_2y_2 Q^2} \phi(y) \nonumber\\
&=&\frac{16\pi \alpha_s(Q^2) f_\pi^2}{Q^2},
\label{Fpias}
\eea
where $[dx]=dx_1 dx_2 \delta(1-x_1-x_2)$, $[dy]=dy_1 dy_2\delta(1-y_1-y_2)$,
$f_\pi= 93$ MeV is the pion decay constant,
and $\phi(x)$ is the distribution amplitude of the pion meson.
The asymptotic form for the distribution amplitude
has been employed in obtaining the send expression in Eq. (\ref{Fpias}),
since any distribution amplitudes for the pion meson
should approach the asymptotic form as $Q^2 \rightarrow \infty$,
\bea
\phi^{(as)}(x)=\sqrt{3}f_\pi x_1x_2.
\label{DAas}
\eea
Eq. (\ref{Fpias}) gives only 1/3 of the experimental data at intermediate
energy region.
Although the Chernyak and Zhitnitsky (CZ) model for the distribution amplitude
\bea
\phi^{(CZ)}(x)=5\sqrt{3}f_\pi x_1x_2(x_1-x_2)^2
\label{DACZ}
\eea
may enhance the prediction for the pion form factor to
the correct direction, the perturbative calculation with CZ distribution
amplitude has been criticized seriously \cite{Isgur, Radyushkin}
because the nonperturbative end-point region is much emphasized
in the CZ model.
Recently studies on the pion-photon transition form factor
\cite{Kroll}
also show that the pion distribution amplitude at currently experimentally
available energy region is much like the asymptotic form but not the CZ form.
Hence, how to match the perturbative calculation with the experimental
data is an interesting issue.
There are two possible explanations:
one is that non-perturbative contributions will dominate in this region;
the other is that non-leading order contributions
in perturbative expansions may be also important in this region.
To make choice between the two possible explanations one needs
to analyze all of the important non-leading contributions carefully.
These contributions come from higher-twist effects,
higher order in $\alpha_s$ and higher Fock states {\it etc.}.
Field, Gupta, Otto and Chang \cite{Field}
pointed out that for the pion form factor the contribution from
the next-leading order in $\alpha_s$ is about $20\%\sim 30\%$.
Employing the
modified hard scattering approach \cite{LiPion,LiProton},
Refs. \cite{JacobK} and \cite{CaoPion2} considered the transverse momentum
effect in the wave function and found that the transverse momentum dependence
in the wave function plays the role to suppress perturbative prediction.
More recently, Tung and Li \cite{Tung} reexamine the perturbative
calculation for the pion form factor in the MHSA
by respecting the evolution of
the pion wave function in $b$ (the transverse extent of the pion)
and employing the two-loop running coupling constant in
the Sudakov form factor.
It is found \cite{Tung} that the evolution of the pion wave function in $b$
improves the match of perturbative prediction with the experimental data.
However, in order to answer the question whether the perturbative calculation
is able to make reliable prediction for the exclusive processes
at currently experimentally available energy region,
the other non-leading contributions such as that from higher twist effects
and higher Fock states \cite{Brodsky,CaoHH} should also be analyzed carefully.

It has been expected that the power corrections to the
pion form factor ($\sim 1/Q^4$) which come form the higher twist terms
of the pion wave function may be important
in the intermediate energy region \cite{CZRep,GesTPLB,GesTSJNP,HuangCS,SzcW}
since there is a large factor
$\sim m_\pi^2/m_0$ ($m_\pi$ being the pion meson mass and $m_0$ being the
mean u- and d-quark masses) in the twist-3 wave function.
However, the calculations for these higher twist contributions are
more difficult than that for the leading twist (twist-2)
because of the end-point singularity becoming more serious.
The leading twist wave functions in the initial and final states being
proportional to $x_1 x_2$ ($x_1$ and $x_2$ being
the fractional momenta carried by the quark and anti-quark)
and $y_1 y_2$ (see Eq. (\ref{DAas})) may cancel the end-point divergent
factor $1/x_2y_2$ coming from the hard-scattering amplitude.
However, the asymptotic behavior of twist-3 wave function
is $x$- ($y$-)independent (see Eq. (\ref{da3})),
which has no help at all to cure the end-point singularity.
In this case, Sudakov form factor is expected to be able to assure
the reasonableness of the perturbative calculation.
Unfortunately, the estimations for the twist-3 contribution in the medium
energy region do not agree with each other
\cite{CZRep,GesTPLB,GesTSJNP,HuangCS,SzcW} (see Fig.~1).
Ref. \cite{GesTPLB} predicates that
\bea
F_\pi (Q^2)
%\stackrel{Q^2\rightarrow\infty}
=
\frac{16\pi \alpha_s(Q^2) f_\pi^2}{Q^2}
\left\{ 1+\frac{m_\pi^4}{Q^2 m_0^2}\alpha_s^{-8/9}(Q^2)
J^2(Q^2) \right\}
\label{FpiG1}
\eea
with
\bea 
J(Q^2)=
\frac{1}{3}
\left[{\rm ln}{\rm ln}(Q^2/\Lambda_{\rm QCD}^2)+ a \right].
\label{J}
\eea
The first and second terms in Eq. (\ref{FpiG1})
correspond to the leading twist (twist-2)
and next-to-leading twist (twist-3) contributions respectively.
%$m_0\simeq 7\, {\rm MeV}$ is the mean u- and b-quark masses.
%$a=1 \sim 2$ is estimated rough from numerical integral, thereby
%Eq. (\ref{FpiG}) is correct up to about $30\%$ accuracy in the medium
%energy region.
In Ref. \cite{GesTPLB} the double logarithmic (DL) corrections are calculated
in the one loop approximation and it is supposed that the sum of
all DL corrections transforms to the exponential function form
(Sudakov form factor).
Hence it is argued that the divergent factor $1/x_2y_2$ at
$x_2(y_2)\rightarrow 0$ is modified by the following way
\bea
\frac{1}{x_2y_2}
\rightarrow
\frac{1}{x_2y_2} \rm{exp}\left\{-[\alpha_s(Q^2)/2\pi]C_F
L(x_2, y_2, k_\perp,l_\perp)\right\},
\eea
with
\bea
L(x_2, y_2, k_\perp,l_\perp)
&=&\rm{ln}(Q^2/k_\perp^2) \rm{ln}(1/x_2) -\rm{ln}^2(1/x_2)
+\rm{ln}(Q^2/l_\perp^2) \rm{ln}(1/y_2) -\rm{ln}^2(1/y_2) \nonumber\\
&&-\rm{ln}(1/x_2)\rm{ln}(1/y_2).
\eea
It is argued \cite{GesTPLB} that the integral with
function $L(x_2, y_2, k_\perp,l_\perp)$
can not be calculated unambiguously. This uncertainty is
incorporated to the factor $a$ being $1\le a\le 2$
in the function $J(Q^2)$ (Eq. (\ref{J})).
According to Eq. (\ref{FpiG1}), the twist-3 contribution is larger
than the asymptotic term (twist-2 contribution) in the region of 
$Q^2\le 30 {\rm \, GeV}^2$.
Ref. \cite{GesTSJNP} includes the Sudakov corrections in a similar
way as Ref. \cite{GesTPLB} but
improved the estimation on the function $J(Q^2)$
and the running mass $m(Q^2)$, and gives
\bea
F_\pi (Q^2)
%\stackrel{Q^2\rightarrow\infty}
=
\frac{16\pi \alpha_s(Q^2) f_\pi^2}{Q^2}
\left[ 1+\frac{m_\pi^4}{Q^2 m_0^2}
\frac{\pi}{6\alpha_s(Q^2)}
\left (\frac{\alpha_s(1 {\rm GeV}^2)}{\alpha_s(Q^2)}\right )^{8/9}
\right].
\label{FpiG2}
\eea
It can be found from Eq. (\ref{FpiG2}) that
the twist-3 contribution is larger that the twist-2 contribution
at about $Q^2\le 15 \, {\rm GeV}^2$.
Ref. \cite{HuangCS} analyzes the Sudakov effects
by introducing an cut-off on the integral region instead of introducing
the transverse momenta $k_\perp$ and $l_\perp$,
and gives another prediction
\bea
F_\pi (Q^2) 
%\stackrel{Q^2\rightarrow\infty}
=
\frac{16\pi\alpha_s(Q^2) f_\pi^2}{Q^2}
\left[1+\frac{m_\pi^4}{Q^2 m_0^2}\frac{1}{6}\left({\rm ln}
\frac{Q^2}{\Lambda_{\rm QCD}^2}\right)^{8/9}\right].
\label{FpiH}
\eea
Eq. (\ref{FpiH}) tells us that the twist-3 contribution
is about $2\sim 0.6$ of the leading twist contribution at the energy region
$2 \, {\rm GeV}^2 \le Q^2 \le 10\,{\rm GeV}^2$.
All of the above calculations (Eqs. (\ref{FpiG1}), (\ref{FpiG2})
and (\ref{FpiH})) give correct power suppression $(\sim 1/Q^2)$ behavior for
the twist-3 contribution in the large $Q^2$ region,
but their predictions for the dependence on
${\rm ln}Q^2$ are very different.
%Thus, Refs. \cite{GesTPLB}, \cite{GesTSJNP} and \cite{HuangCS}
%gave different predictions for the twist-3 contribution in the medium
%energy region.
The main reason for these differences is that
Sudakov corrections are evaluated in different approximations
in Refs. \cite{GesTPLB}, \cite{GesTSJNP} and \cite{HuangCS}.
In the modified hard scattering approach for the exclusive processes
proposed by Li and Sterman \cite{LiPion,LiProton},
the customarily neglected partonic transverse
momentum are combined with Sudakov corrections,  and the Sudakov
form factor is expressed in a more convenient space ($b$-space),
which provides
an more reliable and systematical way to evaluate the Sudakov effect.
Li and Sterman's formalism is originally obtained for
studying the contribution from the leading twist wave function.
%We point out that the MHSA can be extended to evaluate
%the non-leading twist perturbation contributions such as higher twist wave
%function contribution also.
We point out that for the pion electromagnetic form factor
the MHSA can be extended to evaluate
the contribution coming from the twist-3 terms of the pion wave function.
One manifest advantage of MHSA is that there is no other phenomenological
parameter but the input wave function need to be adjusted.
The purpose of this work is to analyze the twist-3 wave function
contribution to the pion form factor in the framework of MHSA.

\vskip 0.5cm
\noindent
{\bf 2 Formalism}
%{\bf 2. Modified Hard Scattering Approach}
\vskip 5pt

\noindent
We first review the derivation of the modified
hard-scattering formalism
for the leading twist (twist-2) contribution to
the pion form factor \cite{LiPion}.
Taking into account
the transverse momenta ${\bf k}_\perp$ and ${\bf l}_\perp$ that 
flow from the wave functions through the hard scattering
leads to a factorization form with two wave
functions $\psi(x,{\bf k}_{\perp})$ and $\psi(y,{\bf l}_{\perp})$
corresponding to the external pions combined with a hard-scattering
function $T_{H}(x,y,Q,{\bf k}_{\perp},{\bf l}_{\perp})$,
which depends in general on transverse as well as longitudinal
momenta,
\begin{eqnarray}
F_{\pi}^{(t=2)}(Q^{2})=\int[dx][dy]\int d^{2}{\bf k}_{\perp} 
d^{2}{\bf l}_\perp \psi(x,{\bf k}_\perp,P_{1},\mu)
T_{H}^{(t=2)}(x,y,Q,{\bf k}_\perp,{\bf l}_\perp,\mu)
\psi(y,{\bf l}_\perp,P_{2},\mu),
\label{fk}
\end{eqnarray}
where
%$[dx]=dx_1 dx_2 \delta(1-x_1-x_2)$, $[dy]=dy_1 dy_2\delta(1-y_1-y_2)$,
$P_1$ and $P_2$ are the momenta of the incoming and outgoing
pion respectively,
$Q^{2}=2P_{1}\cdot P_{2}$ and $\mu$ is the renormalization
and factorization scale.
%In Eq. (\ref{fk}), only the leading twist (twist-2) wave function of pion
%is taken into account.
To the lowest order in perturbation theory,
the hard-scattering amplitude $T_H^{(t=2)}$ is to be calculated from
one-gluon-exchange diagrams.
Neglecting the transverse momentum dependence in the numerator
of $T_H^{(t=2)}$ one can obtain,
\begin{eqnarray}
T_{H}^{(t=2)}(x,y,Q,{\bf k}_\perp,{\bf l}_\perp,\mu)=
\frac{16\pi C_{F} \alpha_{s}(\mu) x_2 Q^2}
{
\left[ x_2 Q^2+{\bf k_\perp}^2\right]
\left[ x_2y_2Q^{2}+({\bf k}_{\perp}-{\bf l}_{\perp})^{2}\right]
},
\label{TH}
\end{eqnarray}
where $C_{F}=4/3$ is the color factor and $\alpha_s(\mu)$ is
the QCD running coupling constant.
The first and the second terms in the denominator come from 
fermion and gluon propagators respectively.

Eq. (\ref{fk}) can be expressed in the ${\bf b}$- and
${\bf h}$-configurations via Fourier transformation
\begin{eqnarray}
F_{\pi}^{(t=2)}(Q^{2})=\int [dx][dy]\frac{d^2{\bf b}}{(2\pi)^{2}}
\frac{d^2{\bf h}}{(2\pi)^{2}}
\varphi(x,{\bf b},P_{1},\mu)
T_{H}^{(t=2)}(x,y,Q,{\bf b},{\bf h},\mu)
\varphi(y,{\bf h},P_{2},\mu),
\label{fb}
\end{eqnarray}
where wave functions $\varphi(x,{\bf b},P_{1},\mu)$
and $\varphi(y,{\bf h},P_{2},\mu)$
take into account an infinite 
summation of higher-order effects associated with the elastic
scattering of the valence partons, which give Sudakov 
suppressions to the large-$b(h)$ and small-$x(y)$
regions \cite{LiPion,CaoPion,BottsS},
\begin{eqnarray}
\varphi(\xi,{\bf b},P,\mu)=\exp\left[ -s(\xi,b,Q)-s(1-\xi,b,Q)
-2 \int_{1/b}^{\mu}\frac{d\bar{\mu}}{\bar{\mu}}\gamma_{q}
(g(\bar{\mu}))\right]\times \phi\left(\xi,\frac{1}{b}\right).
\label{varphi}
\end{eqnarray}
Here $\gamma_{q}=-\alpha_{s}/\pi$ is the quark anomalous dimension.
%in the axial gauge.
$s(\xi,b,Q)$ is Sudakov exponent factor \cite{LiPion,CaoPion,BottsS},
\begin{eqnarray}
s(\xi,b,Q)&=&\frac{A^{(1)}}{2\beta_{1}}\hat{q}
\ln\left(\frac{\hat{q}}{-\hat{b}}\right)+\frac{A^{(2)}}{4\beta_{1}^{2}}
\left(\frac{\hat{q}}{-\hat{b}}-1\right)-\frac{A^{(1)}}{2\beta_{1}}
(\hat{q}+\hat{b})\nonumber \\
& &-\frac{A^{(1)}\beta_{2}}{4\beta_{1}^{3}}\hat{q}
\left[\frac{\ln(-2\hat{b})+1}{-\hat{b}}-\frac{\ln(-2\hat{q})+1}
{-\hat{q}}\right] \nonumber \\
& &-\left(\frac{A^{(2)}}{4\beta_{1}^{2}}-
\frac{A^{(1)}}{4\beta_{1}}\ln(\frac{1}{2}e^{2\gamma-1})\right)
\ln\left(\frac{\hat{q}}{-\hat{b}}\right) \nonumber \\
& &+\frac{A^{(1)}\beta_{2}}{8\beta_{1}^{3}}
\left[\ln^{2}(2\hat{q})-\ln^{2}(-2\hat{b})\right],
\label{smalls}
\end{eqnarray}
where 
\bea
&\hat{q}&=\ln[\xi Q/(\sqrt{2}\Lambda)],~~~~ \hat{b}=\ln(b\lambda),\nonumber\\
&\beta_{1}&=\frac{33-2n_{f}}{12},~~~~
\beta_{2}=\frac{153-19n_{f}}{24},\nonumber\\
&A^{(1)}&=\frac{4}{3},~~~ A^{(2)}=\frac{67}{9}-\frac{1}{3}\pi^{2}
-\frac{10}{27}n_{f}+\frac{8}{3}\beta_{1}\ln(\frac{1}{2}e^{\gamma}).
\eea
$n_{f}$ is the number of quark flavors and $\gamma$ is the Euler constant.
In the derivation of Eq. (\ref{smalls}), the one-loop running coupling
constant has been employed.
It is pointed out \cite{LiBD} that 
additional two terms will appear in the $s(\xi,b,Q)$
expression if the two-loop running coupling constant is used. 
The two terms reduce the prediction for the pion form factor
by only a few percent \cite{LiBD} in the intermediate energy region.
So for simplicity, we neglect these terms.

Applying  the renormalization group equation 
to $T_{H}^{(t=2)}$ and substituting the explicit expression for $T_{H}^{(t=2)}$,
one can obtain the following expression for the pion form factor
\begin{eqnarray}
F_{\pi}^{(t=2)}(Q^{2})&=&\int [dx] [dy]\int b \,db\int h \,dh \,\, 16\pi C_{F}
\alpha_{s}(t) x_2 Q^2 K_{0}(\sqrt{x_2} Qh)
\phi(x,1/b)\phi(y,1/h) \nonumber \\
&~~~\times &\left[\theta(b-h)K_0(\sqrt{x_2} Qb) I_0(\sqrt{y_2}Qh)
       +\theta(h-b)I_0(\sqrt{x_2}Qb) K_0(\sqrt{y_2}Qh)\right]\nonumber \\
&~~~\times & \exp \left(-S(x,y,Q,b,h,t)\right),
\label{Fb2}
\end{eqnarray}
where
\begin{eqnarray}
S(x,y,Q,b,h,t)=
\left[\left(\sum_{i=1}^2 s(x_i,b,Q)+\sum_{i=1}^{2}s(y_i,Q,h) \right)
-\frac{1}{\beta_{1}}\ln\frac{\hat{t}}{-\hat{b}}
-\frac{1}{\beta_{1}}\ln\frac{\hat{t}}{-\hat{h}}
\right].
\label{bigs}
\end{eqnarray}
$K_0$ and $I_0$ are the modified Bessel functions of order zero.
%Radiative corrections in higher orders
%will bring logarithms of the form $\ln(t/\mu)$ into $T_{H}$,
%where $t$ is the largest mass scale appearing in $T_{H}$.
%Ref. \cite{LiPion} points out that a natural choice
%for $\mu$ in $T_{H}$ is $\mu=t$ and
$t$ is the largest mass scale appearing in $T_{H}^{(t=2)}$,
\begin{eqnarray}
t=\max\left(\sqrt{x y} Q,1/b,1/h\right).
\label{Litvalue}
\end{eqnarray}
If $b$ is small, radiative corrections will be small regardless of the values
of $x$ because of the small $\alpha_{s}$. When $b$ is large
and $xyQ^{2}$ is small, radiative corrections are still large in
$T_{H}^{(t=2)}$, but $\varphi$ will suppress these regions.
In Eq. (\ref{Fb2}), $\phi(x,1/b)$ and $\phi(y,1/h)$ are two
input ``wave functions" which respect the non-perturbative physics.
In the large-$Q^2$ region, they can be taken as the asymptotic form
of the twist-2 distribution amplitude (Eq. (\ref{DAas}))
\cite{LiPion,CaoPion,BottsS}.

In the above discussion, only the leading twist wave function is considered.
Now, we address the contributions coming form the twist-3 wave functions.
%Generally, the pion wave function can be written as
%\bea
%\Phi(p,P)=\gamma_5 {\rlap/ P} \Phi_A(p,P)+\gamma_5{\rlap/ p}\Phi_{A'}(p,P)
%          +\gamma_5\Phi_P(p,P)
%          +\gamma_5 P^\mu\sigma_{\mu\nu} p^\nu \Phi_T(p,P),
%\label{PhiBS}
%\eea
%where $\Phi_i(p,P)$ for $i=P,\, A,\, A'$ and $T$ are scalar functions of
%$p^2,\,P^2$ and $(p\cdot P)$ with $p$ and $P$ being
%the relative and total momenta of the system, respectively.
%Here the subscripts $i=P,\, A,\, A'$ and $T$
%denote the spinor matrix structure of the component of $\Phi$.
%Functions $\Phi_A, \,\Phi_P,$ and $\Phi_T$ are even functions of
%$(p\cdot P)$, while $\Phi_{A'}$ is an odd function of $(p\cdot P)$.
%In Eq. (\ref{PhiBS}), the first and the second terms are twist-2 parts,
%while the other two terms belong to twist-3 section.
%The two twist-3 terms might mix under the consideration of the evolution
%equation for two-quark state in the pseudoscalar channel, thus
%the twist-3 wave function of pion can be expressed as \cite{GesTPLB,GesTSJNP},
The operators which contribute to the twist-3 parts of the pion wave function
include $\gamma_5$ and $\gamma_5 \sigma_{\mu\nu}$,
and the two matrixes might mix under the consideration of the evolution
equation for two-quark state in the pseudoscalar channel.
It is pointed out in Refs. \cite{GesTPLB,GesTSJNP}
that the twist-3 wave function of pion can be expressed as
\bea
\psi^{(t=3)}(x,{\bf k}_\bot)\simeq \gamma_5\phi_3
   \left[1- {i} \frac{2(x_1-x_2)}{Q^2}P_1^\mu\sigma_{\mu\nu}P_2^\nu
   -{i} \frac{x_1x_2}{{\bf k}_\perp^2}P_1^\mu\sigma_{\mu\nu}k_\perp^\nu
   \right],
\eea
where ${\bf k}_\perp$ is the partonic transverse momentum.
%According to its definition, we have also
%\be
%          \Phi(p,P)=\int d^4 x \,{\rm e}^{-ip\cdot x}
%	   \langle 0\left| T\left(\psi\left(\frac{x}{2}\right)
%	   {\bar \psi}\left(-\frac{x}{2}\right) \right)\right|P\rangle.
%\ee
%Employing operator production expansion (OPE), we can obtain the
$\phi_3$ is the distribution amplitude of twist-3
\cite{CZRep,GesTPLB,GesTSJNP,HuangCS},
\be
\phi_3=\frac{f_\pi}{4\sqrt{n_c}}\frac{m_\pi^2}{{\bar m}(Q^2)},
\label{da3}
\ee
where $m_\pi=139$ MeV is the pion meson mass and
${\bar m}(Q^2)$ is the mean value of the u- and d-quarks masses
at the scale $Q^2$,
\bea
{\bar m}(Q^2)=\left(\frac{\alpha_s(Q^2)}
{\alpha_s(\mu_0^2)}\right)^{4/\beta_0} m_0(\mu_0^2),
\eea
with $\beta_0=11-\frac{2}{3}n_f$, and
$m_0(1 \, {\rm GeV}^2)=7\pm 2 \,\, {\rm MeV}$.

The hard scattering amplitude for the twist-3 wave function differs from
that for the twist-2 wave function, which turns out to be
\bea
T_H^{(t=3)}(x,y,Q, {\bf k}_\perp,{\bf l}_\perp,\mu)=
\frac{64 \pi C_F \alpha_s(\mu) x_2}
{\left[x_2 Q^2+{\bf k_\perp}^2\right]
\left[ x_2y_2Q^{2}+({\bf k}_{\perp}-{\bf l}_{\perp})^{2}\right]}.
\label{TH3}
\eea
Following the derivation for the leading twist wave function we can obtain
the twist-3 contribution to the pion form factor in the modified 
hard-scattering approach,
\begin{eqnarray}
F_{\pi}^{(t=3)}(Q^{2})&=&\int [dx] [dy]\int b db\int h dh 64\pi C_{F}
\alpha_{s}(t) x_2 K_{0}(\sqrt{x_2} Qh)
\phi_3(x)\phi_3(y) \exp \left(-S(x,y,Q,b,t)\right)\nonumber \\
&\times&\left[\theta(b-h)K_0(\sqrt{x_2}Qb) I_0(\sqrt{y_2}Qh)
       +\theta(h-b)I_0(\sqrt{x_2}Qb) K_0(\sqrt{y_2}Qh)\right].
\label{Fb3}
\end{eqnarray}
It can be found that the hard scattering amplitudes
$T_H^{(t=2)}$ (Eq. \ref{TH})
and $T_H^{(t=3)}$ (Eq. \ref{TH3}) are divergent in the end-point region
$x,y \rightarrow 0, \,\, k_\bot, l_\bot \rightarrow 0$.
However,
the twist-2 contribution to the pion form factor can
be calculated readily because the twist-2 wave functions being
proportional to $x_1 x_2$ and $y_1 y_2$ respectively
cancel the divergent factor $1/x_2 y_2$ in the $T_H^{(t=2)}$.
Furthermore, the Sudakov corrections also suppress the contribution
from the end point region.
For the twist-3 contribution, the wave function is a constant in the
whole region of $x$ (see Eq. (\ref{da3})), which has no help at all
to cure the divergent factor $1/x_2 y_2$ in the $T_H^{(t=3)}$.
In this case, the Sudakov form factor
may guarantee that the calculation is reliable since 
the factor ${\rm e}^{-S}$ rapidly decreases to zero more rapidly than
any power of $x(y)$ at the end-point region (see Eqs. (\ref{smalls})
and (\ref{bigs})).

%It is pointed out in Ref. \cite{JacobK,CaoPion}
%that the b(h)-dependence associated with
%the ``soft" wave functions $\phi(x,1/b)$ and $\phi(y,1/h)$ 
%(intrinsic transverse momentum dependence) suppress
%pQCD predictions for $F_\pi$ in the intermediate energy region also.
%To respect this effect, we adopted the model wave function for 
%$\phi_2(x,1/b)$ in Ref. \cite{JacobK},
%\be
%\phi_2(x,1/b)=A \frac{f_\pi}{2\sqrt{3}}x_1x_2
%{\rm exp}\left[-\frac{\beta^2m_q^2}{x_1x_2}\right]
%{\rm exp}\left[\frac{x_1x_2 b^2}{4\beta^2}\right],
%\ee
%where $A=10.07$ GeV$^{-1}$, $\beta^2=0.883$ GeV$^{-2}$ and $m_q=330$ MeV
%is the constituent quark mass.
%Also we can respect this effect in the twist-3 calculation
%with the following wave function
%\be
%\phi_3(x,1/b)=\frac{f_\pi}{8\sqrt{n_c}}\frac{m_\pi^2}{\bar m}
%\left({\rm ln}\frac{1}{(b\Lambda_{QCD})^2}\right)^{4/(11-2n_f/3)}.
%\ee

\vskip 0.5cm
\noindent
{\bf 3 Numerical result and discussion}
\vskip 5pt

\noindent
We present the numerical evaluations
for the twist-2 and twist-3 contributions to the pion form factor in Fig.~1.
The thinner solid curve is MHSA prediction for the
twist-2 contribution (Eq. (\ref{Fb2})).
The thicker solid curve is twist-3 contribution to the pion form factor
obtained in this work (Eq. (\ref{Fb3})), while
the dash-dotted is the result of Ref. \cite{GesTPLB}
(the second term in Eq. (\ref{FpiG1}) with $a=1.5$).
The dotted and dashed curves are the results given in 
Refs. \cite{GesTSJNP} (the second term in Eq. (\ref{FpiG2}))
and  \cite{HuangCS} (the second term in Eq. (\ref{FpiH}))
respectively.
All of the calculations given in this work, Refs. \cite{GesTPLB},
\cite{GesTSJNP}
and \cite{HuangCS} show that compared with the leading twist contribution,
the twist-3 contributions are suppressed by the factor $1/Q^2$ 
at asymptotic limit ($Q^2 \rightarrow \infty$).
But the predictions are different in the medium and lower energy regions.
Our result is much larger than the result of Ref. \cite{HuangCS}
in the energy region of $2 \, {\rm GeV}^2 \le Q^2 \le 40 \,{\rm GeV}^2$,
and a little larger than the result of Ref. \cite{GesTSJNP}
as $Q^2 \ge 5 \, {\rm GeV}^2$.
Ref. \cite{GesTPLB} and this work
give very similar results in the large-$Q^2$ region,
but our prediction is a little smaller than 
the result of Ref. \cite{GesTPLB} at about  $Q^2 \le 15 \, {\rm GeV}^2$.
The Sudakov corrections are respected systematically in MHSA,
while they are evaluated in various approximations in
Refs. \cite{GesTPLB,GesTSJNP,HuangCS}, so 
the prediction in this work is more reliable.
In Fig.~2, we include both twist-2 and twist-3 contributions
(obtained in this work) to the pion form factor,
and compare with the experimental data.
The dotted and dashed curves are twist-2 and twist-3 contributions
respectively, and the solid curve is the sum.
Compared to the leading (twist-2) contribution, the
twist-3 contribution is negligible at asymptotic limit
since it is suppressed by the factor $1/Q^2$. 
However, the twist-3 contribution is comparable with and even larger than 
the leading twist contribution at intermediate region of $Q^2$
$(2 \, {\rm GeV}^2 \le Q^2 \le 40 \, {\rm GeV}^2)$.
Also it can be found that the perturbative calculations including
both twist-2 and twist-3 contributions are larger than
the experiment data at about $Q^2 \le 5 \, {\rm GeV}^2$.
One can expects that the other nonleading contributions such
as those coming form higher Fock states may be also important
at lower energy regions.

%The numerical prediction for the twist-3 contribution
%given in this work is significantly smaller than that in
%Refs. \cite{GesTPLB,GesTSJNP,SzcW,HuangCS} because of the suppression coming from 
%Sudakov correction and intrinsic transverse momentum dependence.
%The ratio of the twist-3 to the twist-2
%contributions in this work is a little smaller than that
%in Refs. \cite{GesTPLB,GesTSJNP,SzcW,HuangCS}.
%For example, at $Q^2=5$ GeV$^2$ the
%ratios are about $2.3$, $1.3$, $1.25$ and $1.4$ respectively in this work,
%Refs. \cite{GesTPLB,GesTSJNP}, \cite{SzcW} and \cite{HuangCS}.
%This is because our result is obtained
%in the MHSA while Refs. \cite{GesTPLB,GesTSJNP,SzcW,HuangCS} work in the standard
%hard-scattering approach.
%Also, it can be found that pQCD calculation
%including both twist-2 and twist-3 contributions
%is consistent with the experimental data in the higher energy region.
%Therefor we conclude that pQCD calculation including both leading 
%contribution and power corrections is able to make reliable prediction
%for the exclusive processes in the $Q^2$ region where currently
%experimental data are available.

In summary, we analyzed the twist-3 contribution to
the pion electromagnetic form factor in the modified
hard scattering approach in which Sudakov corrections are
respected systematically, and compared with various approximate
calculations.
It is found that the twist-3 contribution
is enhanced significantly since
the twist-3 wave function is independent of the
fractional momentum carried by the parton
and has a large factor $\sim m_\pi^2/m_0$, while the twist-2
wave function is proportional to $x_1 x_2$ ($y_1 y_2$) which cancels the
end-point divergent factor $1/x_2y_2$ in the hard-scattering amplitude.
Thus, although it is suppressed by the factor $1/Q^2$
as compared with the leading (twist-2) contribution,
the twist-3 contribution is comparable with and even large than
the leading twist contribution at intermediate region of $Q^2$
being $2 \sim 40 ~{\rm GeV}^2$.
The perturbative predictions including both twist-2 and twist-3
contributions are larger than the experiment data at lower
energy regions, which indicates the importance to
study the other nonleading corrections at these energy regions.

%Also it is found that
%the pQCD prediction including both twist-2 and twist-3 contributions
%is consistent with the experimental data in the higher energy region.
%Therefor we conclude that pQCD calculation including both leading 
%contribution and power corrections is able to make reliable prediction
%for the exclusive processes
%in the $Q^2$ region where currently experimental data are available.

\vskip 0.5cm
\noindent
{\it Acknowledgments.}
This work partially supported by the
Postdoc Science Foundation of China and the National Natural
Science Foundation of China.
F. G. Cao would like to thank professor B. V. Geshkenbein for providing
more inferences for their work \cite{GesTPLB},
and professor M. V. Terentyev for providing the preprint/ITEP-45 (1982).

%\newpage
\vskip 1cm
\noindent
{\bf Figure caption}
\begin{description}
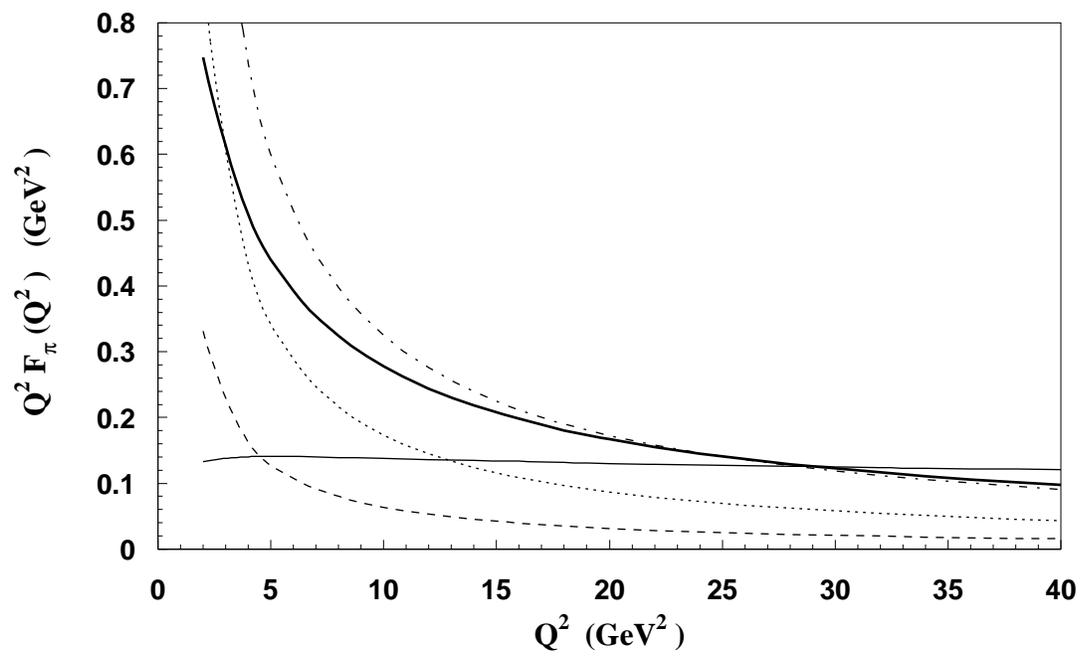
\item{Fig.~1}
Twist-2 and twist-3 contributions to the pion form factor.
Each curve is explained in the text.
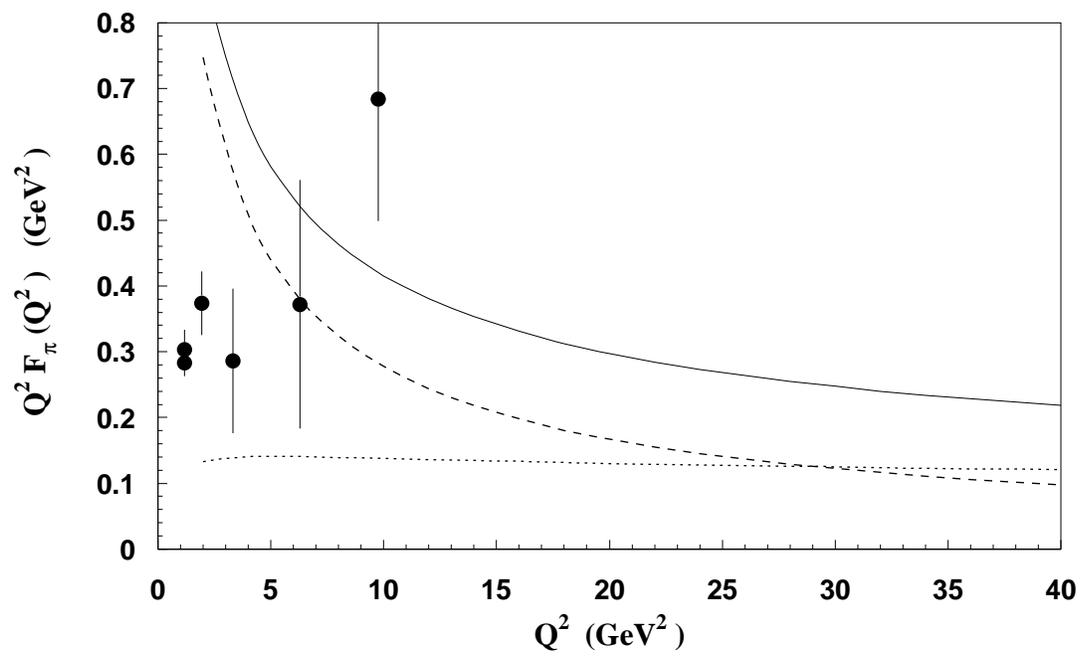
\item{Fig.~2}
Perturbative prediction for the pion form factor including both
twist-2 (dotted curve) and twist-3 (dashed curve) contributions.
The solid curve is the sum of twist-2 and twist-3 contributions.
The data are taken from Ref. \cite{Data}.
\end{description}
%}
\end{document}